\documentclass[%
    reprint,
    preprintnumbers,
    superscriptaddress,
    amsmath, amssymb,aps,prc,
    floatfix,
    nolongbibliography
]{revtex4-2}

\usepackage{palatino}
\usepackage{mathpazo}
\usepackage{graphicx}
\graphicspath{{figures/}}
\usepackage{color}
\usepackage[breaklinks,colorlinks,urlcolor=blue,citecolor=blue,linkcolor=blue]{hyperref}

\newcommand{\pp}{\ensuremath{p}+\ensuremath{p}}
\newcommand{\PbPb}{Pb+Pb} 
\newcommand{\AAcoll}{A+A}

\newcommand{\GeVc}{\ensuremath{\mathrm{GeV}\kern-0.05em/\kern-0.02em c}}
\newcommand{\GeV}{\ensuremath{\mathrm{GeV}}}
\newcommand{\TeV}{\ensuremath{\mathrm{TeV}}}
\newcommand{\pT}{\ensuremath{p_{\mathrm{T}}}}

\newcommand{\pTjet}{\ensuremath{p_{\mathrm{T}}^{\mathrm{jet}}}}
\newcommand{\antikt}{\ensuremath{\mathrm{anti\mbox{-}}k_t}}

\newcommand{\DeltaR}{\ensuremath{\Delta R_{\mathrm{axis}}}}
\newcommand{\DeltaRWtaStd}{\ensuremath{\Delta R_{\mathrm{axis}}^{\mathrm{WTA}\text{\textendash}\mathrm{Std}}}}
\newcommand{\snn}[2]{\ifmmode{\sqrt{s}=#1~\mathrm{#2}\ignorespaces}\else%
    \mbox{$\sqrt{s}=#1~\mathrm{#2}$}\ignorespaces\fi
}

\newcommand{\PYTHIA}{\textsc{Pythia}}
\newcommand{\FASTJET}{\textsc{FastJet}}
\newcommand{\TRENTo}{\textsc{TrENTo}}
\newcommand{\MUSIC}{\textsc{Music}}
\newcommand{\LBT}{\textsc{Lbt}}

\begin{document}

\title{Phenomenological study of the angle between jet axes in heavy-ion collisions}


\author{Jin-Wen Kang}
\affiliation{
	Key Laboratory of Quark \& Lepton Physics (MOE) and Institute of Particle Physics,
	Central China Normal University, Wuhan 430079, China
}

\author{Sa Wang}
\affiliation{College of Science, China Three Gorges University, Yichang 443002, China}
\affiliation{Center for Astronomy \& Space Sciences and Institute of Modern Physics, China Three Gorges University, Yichang 443002, China}

\author{Lei Wang}
\affiliation{%
	School of Applied Physics and Materials, Wuyi University, Jiangmen 529020, China
}

\author{Ben-Wei Zhang}
\email{bwzhang@mail.ccnu.edu.cn}
\affiliation{
	Key Laboratory of Quark \& Lepton Physics (MOE) and Institute of Particle Physics,
	Central China Normal University, Wuhan 430079, China
}

\date{\today} 

\begin{abstract}
	This paper presents a phenomenological study on the angle between the Standard and 
	Winner-Take-All (WTA) jet axes (\(\Delta
	R_{\mathrm{axis}}^{\mathrm{WTA}\text{\textendash}\mathrm{Std}}\)) in high-energy nuclear
	collisions. The $p$+$p$ baseline is provided by the \textsc{Pythia}8 event generator. The in-medium
	jet propagation is simulated by the linear Boltzmann transport (\textsc{Lbt}) model, which
	considers both the elastic and inelastic jet-medium interactions. Our theoretical results
	calculated by the \textsc{Lbt} model show that the \(\Delta
	R_{\mathrm{axis}}^{\mathrm{WTA}\text{\textendash}\mathrm{Std}}\) distribution in Pb+Pb at
	\(\sqrt{s}=5.02~\mathrm{TeV}\) is narrower than that in $p$+$p$, which agrees well with the recent
	ALICE measurements. The narrowing of \(\Delta
	R_{\mathrm{axis}}^{\mathrm{WTA}\text{\textendash}\mathrm{Std}}\), which seems to violate the nature of intra-jet broadening due to jet quenching, may be attributed to the influence of ``selection bias''.
    However, the physical details still need to be fully understood. Utilizing a
	matching-jet method to track the jet evolution in the QGP to remove the selection bias in the Monte
	Carlo simulations, we observe that the \(\Delta
	R_{\mathrm{axis}}^{\mathrm{WTA}\text{\textendash}\mathrm{Std}}\) distribution becomes broader due
	to the jet-medium interactions. At the same time, by rescaling the quark/gluon-jet fractions in
	Pb+Pb collisions to be the same as that in $p$+$p$, we find that the fraction change may not
	significantly influence the modification pattern of jet \(\Delta
	R_{\mathrm{axis}}^{\mathrm{WTA}\text{\textendash}\mathrm{Std}}\). On the other hand, the selected
	jet sample in A+A collisions has a significantly narrower initial \(\Delta
	R_{\mathrm{axis}}^{\mathrm{WTA}\text{\textendash}\mathrm{Std}}\) distribution than the $p$+$p$
	baseline, and such a biased comparison between $p$+$p$ and A+A conceals the actual intra-jet broadening
	effect in the experimental measurements. The investigations presented in this paper will deepen our
	understanding of the relationship between the actual intra-jet modifications in the QGP and the
	experimental observations.
\end{abstract}

\maketitle

\section{Introduction}
Exploring the properties of the quark-gluon plasma (QGP), a new state of nuclear matter formed at
extremely hot and dense conditions, is one of the most important motivations of the heavy-ion
collision program at the RHIC and the LHC~\cite{Freedman:1976ub, Shuryak:1977ut, Connors:2017ptx,
	Blaizot:2015lma, Braun-Munzinger:2015hba, Cunqueiro:2021wls, Cao:2022odi, Apolinario:2022vzg}. The
high \pT{} parton or jets generated at a very early stage of the collisions will strongly interact
with the medium and dissipate energy when traversing through the QGP, called the ``jet quenching''
phenomenon. Therefore, jets are widely regarded to be an excellent probe for unveiling the mystery
of such deconfined quark soup~\cite{Wang:1992qdg, Gyulassy:2003mc, Mehtar-Tani:2013pia,
	Qin:2015srf, Wang:1998bha, Wang:2001cs, Wang:2002ri, Vitev:2009rd, Neufeld:2010fj, He:2020iow,
	ALICE:2023dwg, Kang:2023qxb, Dai:2012am, Yang:2023dwc, Wang:2023eer, Xie:2022ght, Yang:2022nei}. A
series of recently developed jet substructure observables have been used to perform the
investigations for further understanding the QGP medium and nucleon structure, including the scaled
groomed jet radius $\theta_g$, the groomed jet momentum splitting fraction
$z_g$~\cite{Ringer:2019rfk,ALICE:2019ykw,ALargeIonColliderExperiment:2021mqf,Wang:2022yrp,ATLAS:2022vii,Zhang:2023jpe,JETSCAPE:2023hqn},
the number of ``Soft Drop'' branches $n_{\mathrm{SD}}$~\cite{ALICE:2019ykw}, the radial
profile~\cite{Li:2022tcr, Wang:2020ukj, Wang:2020bqz, Wang:2019xey}, the generalised angularities
$\lambda_{\beta}^{\kappa}$~\cite{Ehlers:2022dfp}, the jet charge
$Q^{\kappa}$~\cite{Chen:2019gqo,CMS:2020plq}, the groomed jet mass~\cite{CMS:2018fof}, the
splitting parameter $\sqrt{d_{12}}$, the angular separation $\Delta R_{12}$~\cite{ATLAS:2023hso},
the $N$-subjettiness, and the sub-jet fragmentation~\cite{Mulligan:2021sxo}, etc.

Besides the observables mentioned above, there is a fascinating observable---the angle between
differently defined jet axes, $\DeltaR$, which probes a vast phase space of the jet formation and
evolution~\cite{ALICE:2022rdg}. The concept of the angle between jet axes $\DeltaR$ was firstly
proposed in Ref.~\cite{Cal:2019gxa}, in which $\DeltaR$ was calculated at next-to-leading
logarithmic accuracy and consistent with the simulations of the Monte Carlo event generators.
Subsequently, the ALICE collaboration reported experiment measurements of the angle between jet
axes in \pp{} and \PbPb{} collisions at \snn{5.02}{TeV}~\cite{ALICE:2022rdg,ALICE:2023dwg}.

The existing studies show that the physical meanings of the \DeltaR{} observables are clear and
straightforward. \DeltaR{} measures the geometric distance of two different axes determined by two
different algorithms for the same reconstructed jet in the $(\eta, \phi)$ plane. Various algorithms
determine different jet axes, each with varying degrees of sensitivity to the soft radiation. More
specifically, the ``Soft Drop'' grooming algorithm removes the soft wide-angle radiation in the
jets, and the degree of this grooming can be adjusted by parameters~\cite{Larkoski:2014wba}. The
WTA axis is obtained by reclustering using the ``Winner-Take-All'' recombination scheme, which
tracks the energetic collinear radiation. The WTA axis is typically aligned with the most energetic
components of the jet, and the effect of soft radiation is power suppressed in the WTA
axis~\cite{ALICE:2022rdg,Cal:2019gxa}. Systematic discussions of \DeltaR{} in \pp\ collisions are
performed in Ref.~\cite{Cal:2019gxa}.

In nucleus-nucleus collisions, due to the different sensitivity to the soft radiation for different
definitions of the jet axis, the \DeltaR{} provides a unique opportunity to gain insight into the
energy loss mechanisms of jet-medium interactions. Recently the ALICE collaboration has reported
the first measurement on the \DeltaR{} distributions of charged-particle jets in \PbPb\ collisions
at $\sqrt{s_{NN}}=5.02~\TeV$~\cite{ALICE:2023dwg}. It is found that \DeltaR{} has narrower
distributions in \PbPb\ collisions than in \pp. The experimental results show different
modification patterns compared to the theoretical calculations with an intra-jet
broadening~\cite{Ringer:2019rfk}. In Ref.~\cite{ALICE:2023dwg}, it is implied that the
``selection bias'' may play an essential role during the jet finding in A+A collisions. In other
words, the minimum \pTjet{} threshold biases the jet selection to favor the candidates with
less energy loss~\cite{Renk:2012ve,Cunqueiro:2021wls}, but the details in physics are unclear.
Moreover, due to the different Casimir color factors ($C_F=4/3$ for quark and $C_A=3$ for gluon),
the gluon-initiated jets will lose more energy than quark-initiated jets during medium-induced
interactions, which leads to a reduced gluon-jet fraction of jet sample in A+A
collisions~\cite{Chen:2019gqo,Li:2019dre,CMS:2020plq}. 
Since the gluon-jet is expected to interact more with the QGP medium, Ref.~\cite{ALICE:2023dwg} suggests that the reduced gluon-jet fraction (from an alternative perspective, a relatively enhanced quark-jet fraction) is believed to be the main reason for the narrower \DeltaR{} distribution in \PbPb{} collisions compared to the vacuum case.
Further detailed
theoretical explorations are necessary to disentangle the influences of the selection bias and the
quark/gluon-jet fraction changes to the observation in the experiment, which is essential to
understand the relationship between the actual intra-jet modifications in the QGP and the
experimental observations.

This paper presents a theoretical study on the angle between the Standard and WTA jet axes
(\DeltaRWtaStd{}) in high-energy nuclear collisions. The \pp\ baseline is provided by the
\PYTHIA{8} event generator, and the in-medium jet propagation is simulated by the linear Boltzmann
transport (\LBT) model. We will perform the calculations of the \DeltaRWtaStd{} distribution in
Pb+Pb at \snn{5.02}{TeV}\ compared to the recently reported ALICE measurements. Furthermore, we
will investigate the influences of the event ``selection bias'' and the quark/gluon-jet fraction
changes to the modification patterns of \DeltaRWtaStd{} distribution in nucleus-nucleus collisions.
We will use the ``matching-jet'' method to track the jet evolution in the QGP, in which the
realistic jet modification can be unveiled. In addition, we will perform a detailed analysis to
demonstrate that the critical factor concealing such intra-jet broadening effects in experimental
measurement is the biased comparison between two jet samples in \pp\ and \PbPb\ collisions. At the
same time, the influence of the quark/gluon-jet fraction changes is limited.

The rest of this article is structured as follows. In Sec.~\ref{sec:framework}, we introduce the
theoretical framework utilized to study the medium modification of \DeltaRWtaStd{} in high-energy
nuclear collisions. We show our calculated results and discussions in Sec.~\ref{sec:res}. At last,
we summarize this work in Sec.~\ref{sec:sum}.

\section{Framework}
\label{sec:framework}

When employing different reclustering/grooming algorithms to process a given jet, we will get some
new jets slightly different from the original one. The \DeltaR{} is a novel infrared and collinear (IRC) safe jet-substructure observable
proposed in Ref.~\cite{Cal:2019gxa} to study the angular distance between the two jet axes defined
with different jet algorithms:
\begin{equation}
	\Delta R_{\rm axis}^{a-b}=\sqrt{(y_{\rm axis}^a-y_{\rm axis}^b)^2+(\phi_{\rm axis}^a-\phi_{\rm axis}^b)^2},
\end{equation}
where $a$ and $b$ denote any two different axis algorithms, $y$ and $\phi$ are the rapidity and the azimuthal angle.
According to different reclustering/grooming algorithms, three types of \DeltaR\ combination are extensively
investigated in jet physics at the LHC~\cite{Cal:2019gxa,ALICE:2022rdg,Ehlers:2022dfp,ALICE:2023dwg}:
WTA\,--\,Standard, WTA\,--\,SoftDrop and Standard\,--\,SoftDrop. This work will focus on the angle between
the Standard and WTA axes.

At the LHC, almost universally employed the \antikt{} algorithm~\cite{Cacciari:2008gp} to
reconstruct jets~\cite{Gauld:2022lem}, so we call the jet axis obtained using the \antikt{}
algorithm (with default $E$ recombination scheme) clustering as the ``Standard'' axis. The WTA is a
recombination scheme that specifies how to combine the momenta when merging two subbranches during
the clustering procedure. Specifically, we directly assign the transverse momentum sum of the two
subbranches as the transverse momentum of the merged branch and designate the direction of the
harder of two subbranches as the direction of the merged
branch~\cite{ALICE:2022rdg,Bertolini:2013iqa},
\begin{align*}
	p_{T,r} & =p_{T,i} + p_{T,j},                                            \\
	\phi_r  & =\phi_k,                                                       \\
	y_r     & =y_k,                                                          \\
	k       & =\begin{cases}
		           i & p_{\mathrm{T},i} > p_{T,j} \\
		           j & p_{\mathrm{T},i} < p_{T,j} \\
	           \end{cases},
\end{align*}
where the subscript $r$ denotes the merged branch; $i$ and $j$ denote the two prongs, respectively.
The WTA scheme is infrared/collinear safe and can powerfully suppress the effect of soft
radiation~\cite{Cal:2019gxa,ALICE:2022rdg}. Because of these features, the WTA scheme has been
used to study jet-like event shapes~\cite{Bertolini:2013iqa}, recoil-free
observables~\cite{Larkoski:2014uqa,Neill:2018wtk}, flavor identification of jets~\cite{Caletti:2022glq},
etc. Back to our work, we recluster the constituents of the ``Standard'' jet using the C/A
algorithm~\cite{Dokshitzer:1997in,Wobisch:1998wt} with the WTA \pT{} recombination scheme,
and then we can get the WTA axis.

A Monte Carlo event generator \PYTHIA{} v8.309~\cite{Bierlich:2022pfr} with Monash
2013~\cite{Skands:2014pea} tune is used to simulate jet productions in \pp\ collisions as the
baseline. We use charged-particles with $\pT > 0.15~\GeVc$ and pseudorapidity range $|\eta| < 0.9$
to reconstruct jets. All jets are reconstructed by the \antikt{} algorithm with the $E$
recombination scheme and the distance parameter $R=0.2$ using \FASTJET{} v3.4.0
package~\cite{Cacciari:2005hq,Cacciari:2011ma}.

We employ the linear Boltzmann transport (\LBT{})
model~\cite{Wang:2013cia,He:2015pra,Cao:2016gvr,Cao:2017hhk} to simulate both elastic and inelastic
scattering processes of the jet shower partons and the thermal recoil partons in the QGP medium.
The \LBT\ model performed well in a series of jet quenching measurements, which was highly
consistent with the experimental
results~\cite{Wang:2013cia,Cao:2016gvr,Cao:2017hhk,Luo:2018pto,Zhang:2018urd}. In the \LBT{} model,
the linear Boltzmann transport equation is used to describe the elastic scattering process,
\begin{eqnarray}
	p_1\cdot \partial f_1(p_1) &=& -\int\frac{d^3p_2}{(2\pi)^32E_2}
	\int\frac{d^3p_3}{(2\pi)^32E_3}\int\frac{d^3p_4}{(2\pi)^32E_4} \nonumber\\
	&&\times\frac{1}{2}\sum_{2(3,4)}\left(f_1f_2-f_3f_4\right)\left|\mathcal{M}_{12\to 34}\right|^2
	\left(2\pi\right)^4\nonumber\\ &&\times S_2(\hat{s}, \hat{t},
	\hat{u})\delta^{(4)}(p_1+p_2-p_3-p_4),
\end{eqnarray}
where $f_i(p_i)(i=1,2,3,4)$ are the phase-space distribution of partons that participate in the
reaction; $\left|\mathcal{M}_{12\to 34}\right|^2$ is the leading-order (LO) elastic scattering
matrix elements~\cite{Combridge:1978kx,Eichten:1984eu}; in order to avoid possible collinear
divergence of $\left|\mathcal{M}_{12\to 34}\right|^2$ for massless partons, here, a
Lorentz-invariant regularization condition
$S_2(\hat{s}, \hat{t}, \hat{u})\equiv \theta\left(\hat{s}\geqslant 2\mu_D^2\right)
	\theta\left(-\hat{s}+\mu_D^2\leqslant \hat{t} \leqslant -\mu_D^2\right)$ is introduced~\cite{Auvinen:2009qm};
$\mu_D^2$ is the Debye screening mass. It should be noted that the medium response effect is included in
the \LBT\ model during our simulations. The higher-twist
approach~\cite{Guo:2000nz,Wang:2001ifa,Zhang:2003yn,Zhang:2003wk,Zhang:2004qm} describes the
inelastic scattering of medium-induced gluon radiation,
\begin{equation}
	\frac{dN_g}{dxdk_{\perp}^2dt}=\frac{2\alpha_sC_AP(x)\hat{q}}{\pi k_{\perp}^4}
	\left(\frac{k_{\perp}^2}{k_{\perp}^2+x^2m^2}\right)^4\sin^2\left(
	\frac{t-t_i}{2\tau_f}\right),
\end{equation}
where $x$ is the energy fraction of the emitted gluon taken from its parent parton with mass
$m$, $k_\perp$ is the transverse momentum of the radiated gluon, $P(x)$ is the splitting
function, $\hat{q}$ is the jet transport coefficient (specific details can be found in
Ref~\cite{He:2015pra}), and $\tau_f$ is the formation time of the radiated gluons in the
QGP medium. The background hydrodynamic profile for the bulk QGP medium is provided by
simulations from the 3$+$1D \MUSIC{} relativistic second-order viscous hydro
model~\cite{Schenke:2010nt,Schenke:2010rr,Paquet:2015lta} with the averaged initial
condition from the 3D \TRENTo{} model~\cite{Moreland:2014oya,Ke:2016jrd}. 
The evolution of the background medium was carried out in a few key steps. First, we employed the 3D \TRENTo\ model to generate about 2000 event-by-event 
initial conditions for the most central 0-10\% of \PbPb\ collisions at $\sqrt{s_{NN}}=5.02$ TeV. Next, we averaged these initial conditions to obtain the average initial state profile. 
Ignoring the evolution of the pre-equilibrium stage, we directly used the averaged initial-state condition as the input for the 3+1D \MUSIC\ model to simulate the evolution of the QGP medium.
Consequently, our current framework does not include event-by-event 
fluctuations~\cite{Zhang:2012ik} or pre-equilibrium effects. 

After the evolutions of the parton shower in the medium are completed, we need to perform
hadronization to obtain the hadron-level events that can be observed in the experiment. In this
work, we first construct strings using the colorless method from the \textsc{JetScape} framework
based on minimization criteria~\cite{Putschke:2019yrg} and then perform hadronization and hadron
decays using the Lund string model~\cite{Bierlich:2022pfr,Andersson:1983jt,Sjostrand:1984ic}
provided by \PYTHIA{}. The \textsc{JetScape} collaboration has already demonstrated in
works~\cite{JETSCAPE:2022jer,JETSCAPE:2023hqn} that this model is reliable.

\section{Results and Discussions}
\label{sec:res}

\begin{figure}
	\centering
	\includegraphics[width=1\columnwidth]{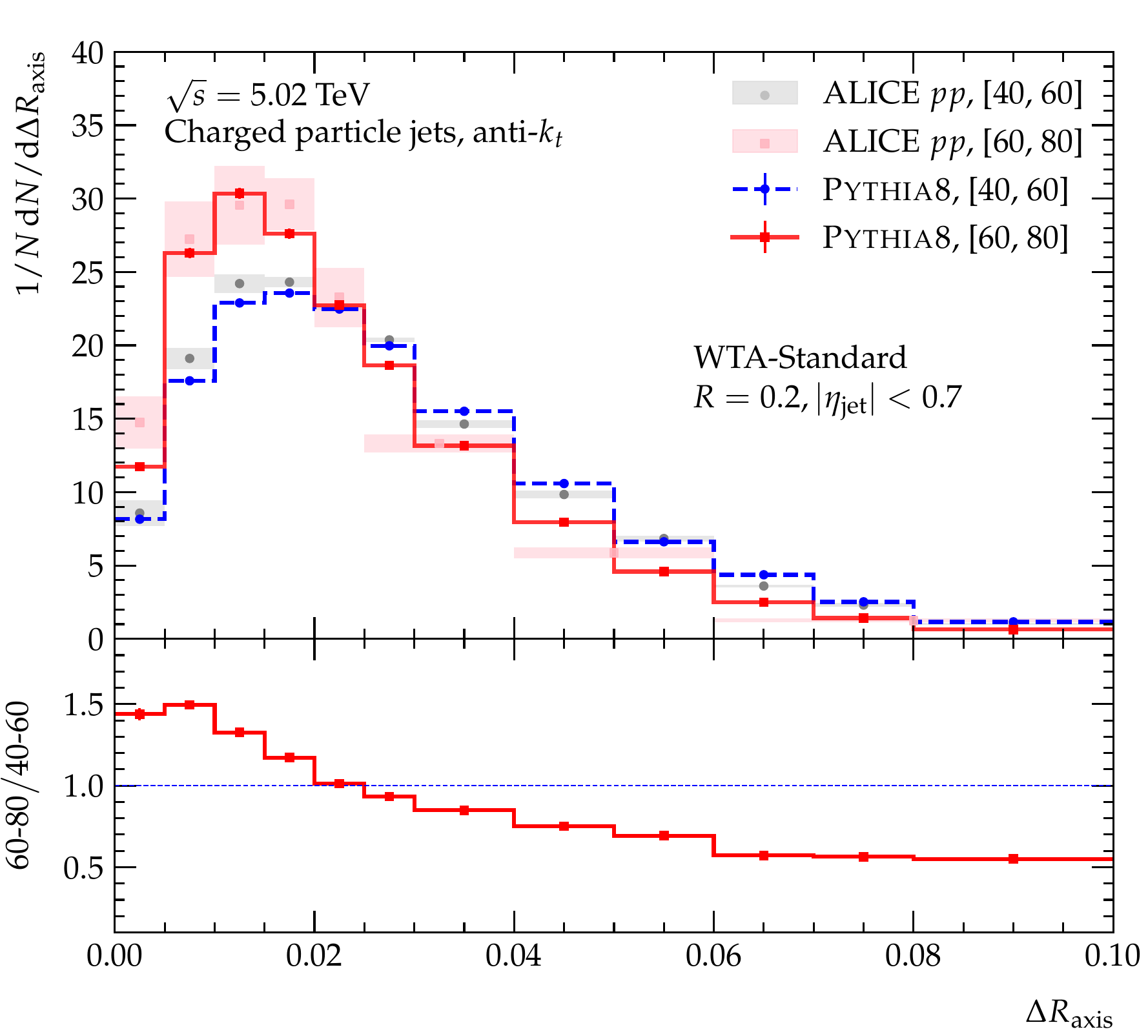}
	\caption{Normalized distributions of the angle between the Standard and WTA
		axes \DeltaRWtaStd{} of inclusive jets calculated by the \PYTHIA{} in \pp\
		collisions at $\sqrt{s_{NN}}=5.02~\TeV$ within two \pT{} ranges: 
		$40<\pT<60~\GeVc$ and $60<\pT<80~\GeVc$, are compared to the ALICE
		data~\cite{ALICE:2023dwg}. The ratio of these two \pT{} ranges
		is also plotted in the lower panel.}
	\label{fig:pp-baseline}
\end{figure}

To calibrate the \pp\ baseline of our study, in Figure~\ref{fig:pp-baseline}, we firstly show the
normalized distributions of the angle between the Standard and WTA axes \DeltaRWtaStd{} of
inclusive jets calculated by the \PYTHIA{} in \pp\ collisions at $\sqrt{s}=5.02~\TeV$ within
two \pT{} ranges $40<\pT<60~\GeVc$ and $60<\pT<80~\GeVc$ compared to the ALICE data. It is found
that the calculations of \PYTHIA{} can describe the ALICE data for these two \pT{} ranges. By taking the
ratio of these two \pT{} ranges in the lower panel, we find that jets with $60<\pT<80~\GeVc$ have
a narrower \DeltaRWtaStd{} distribution than those with $40<\pT<60~\GeVc$, which indicates that
higher-\pT{} jets usually have a narrower \DeltaRWtaStd{} than the lower-\pT{} ones.

\begin{figure}
	\centering
	\includegraphics[width=1\columnwidth]{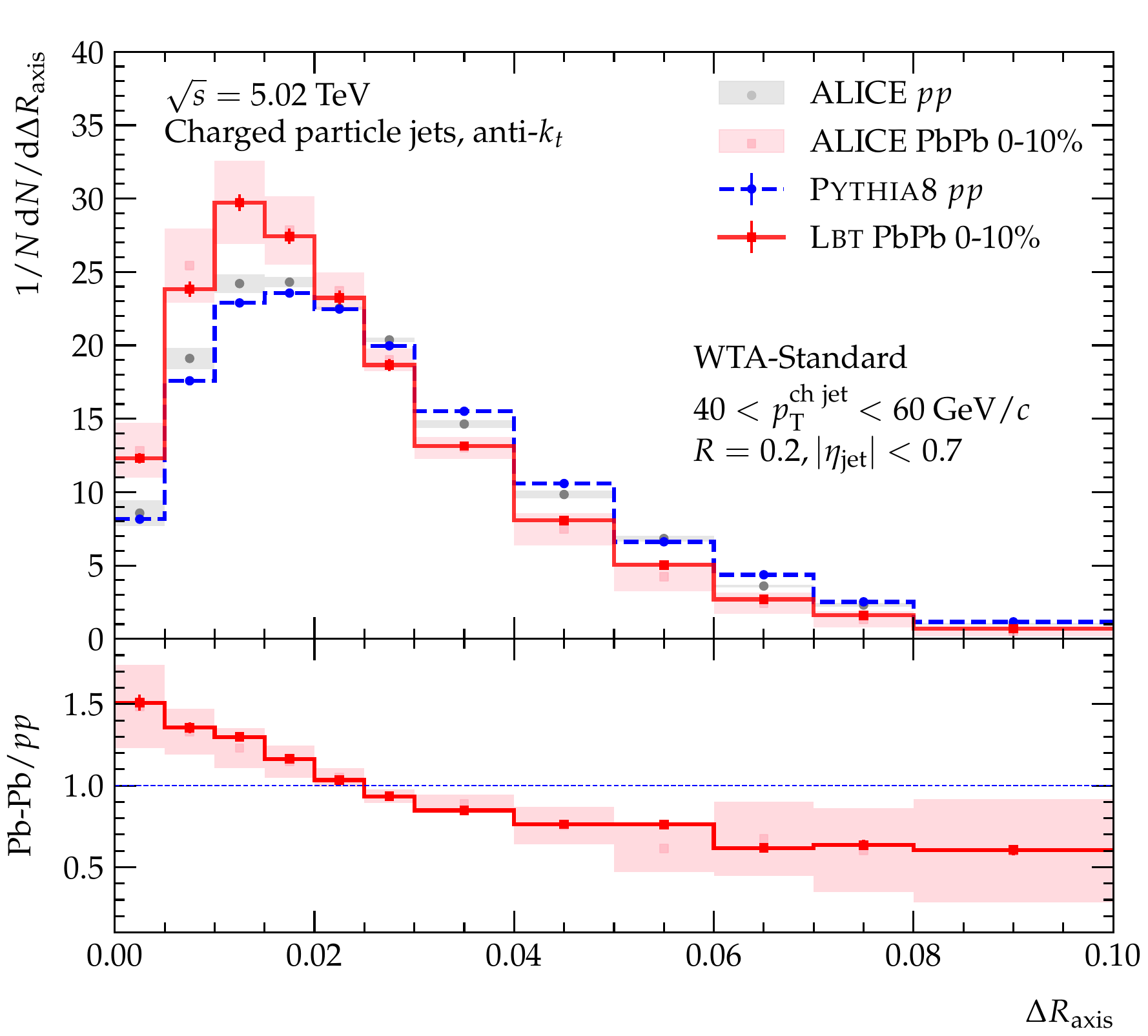}
	\caption{Normalized \DeltaRWtaStd{} distributions for inclusive jets
		in \pp{} and \PbPb{} collisions at $\sqrt{s_{NN}}=5.02~\TeV$, including the results of theoretical
		calculation and ALICE experimental data~\cite{ALICE:2023dwg}. The ratios are also plotted
		in the lower panel.
	}
	\label{fig:model_vs_experimental}
\end{figure}

Now we proceed to the calculations of \DeltaRWtaStd{} distribution in nucleus-nucleus collisions.
In Figure~\ref{fig:model_vs_experimental}, we show the normalized \DeltaRWtaStd{} distribution in
\pp\ and 0-10\% \PbPb\ collisions at $\sqrt{s_{NN}}=5.02~\TeV$ compared to the recent reported ALICE data~\cite{ALICE:2023dwg}.
At the same time, the ratios of PbPb/$pp$ are also shown in the lower panel. Our theoretical
results can reasonably describe the experimental data in both \pp{} and \PbPb{} collisions and the
PbPb/$pp$ ratio. We observe that the \DeltaRWtaStd{} distribution in \PbPb\ collisions is narrower
than that in \pp\, which indicates the two jet axes are closer for the jet sample in \PbPb\
compared to that in \pp.

\begin{figure}
	\centering
	\includegraphics[width=1\columnwidth]{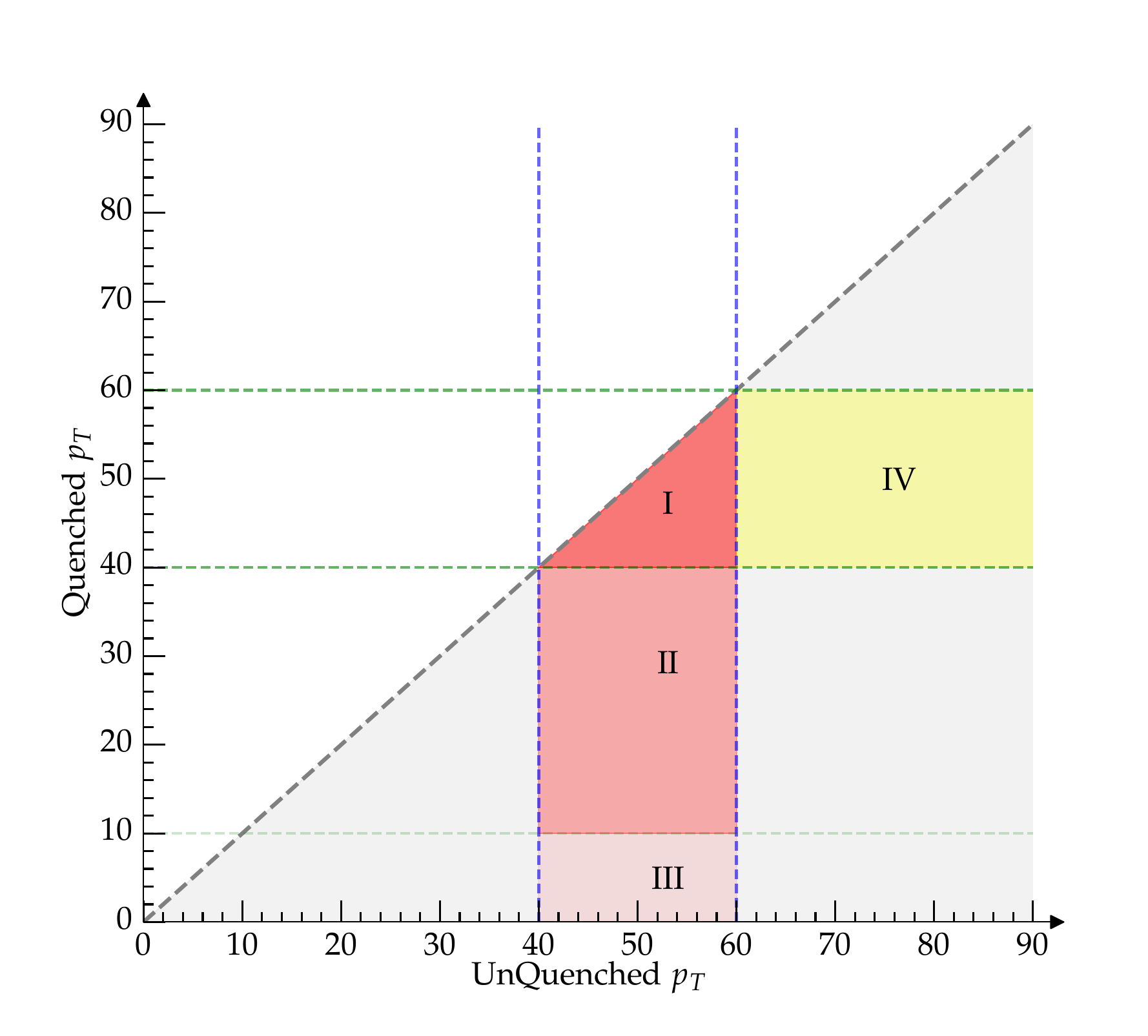}
	\caption{Schematic of the selection bias. The horizontal axis corresponds to the transverse
		momentum of the unquenched jets, and the vertical axis corresponds to the quenched jets' transverse
		momentum. Effective jets exist only in the shaded area of the lower triangle.}
	\label{fig:pt_selection_bias}
\end{figure}

Generally, the same minimum and maximum \pT\ thresholds (e.g. $40 < \pTjet < 60~\GeVc$) are
imposed to select the jet samples both in \pp\ and \AAcoll\ collisions. The comparison of these two
jet samples provides valuable information about the in-medium jet modification relative to the
vacuum case. However, if only considering the jet-medium interaction but not the ``selection
biases'' in the theoretical calculations and experimental measurements, we cannot fully understand
the relationship between the energy loss mechanisms and the observables~\cite{Renk:2012ve}.
Consider a jet (unquenched jet) generated in the initial hard scattering, passing through the QGP
medium and losing energy due to interaction with the medium, becoming a medium-modified jet
(quenched jet). According to the transverse momentum values before and after the energy loss of
this jet, we can place this jet in the lower triangle shadow area in
Figure~\ref{fig:pt_selection_bias}. In usual experimental and phenomenological studies of the
current, the selected sample of \pp{} events, which is the unquenched jets sample, is located in
the I$+$II$+$III region in Figure~\ref{fig:pt_selection_bias}. The selected jet sample in heavy-ion
collisions, the quenched jets, is in the I$+$IV region in Figure~\ref{fig:pt_selection_bias}. This
means the traditional selection method of unquenched and quenched jet samples is significantly
biased. We note that the area I corresponds to the jets survived in the selection window
(40--60~\GeVc) after in-medium energy loss. Moreover, area IV corresponds to jets with higher \pT{}
($>60~\GeVc$) but finally falls in the selection window due to the evident energy loss.

The selection bias effect often plays an important role in the studies of nuclear modification in
heavy-ion collisions. One can find the related investigations in
Refs.~\cite{Baier:2001yt,Lai:2021ckt,Renk:2012ve,Brewer:2018mpk,Casalderrey-Solana:2018wrw,Brewer:2020chg,Du:2020pmp,Brewer:2021hmh,Cunqueiro:2021wls,Wang:2020ukj}.
We have noticed that there have been a lot of research (e.g.~\cite{Brewer:2018dfs,Brewer:2020chg,Du:2020pmp,Takacs:2021bpv,Brewer:2021hmh,Wang:2021jgm,ALICE:2023qve,ALICE:2023jye,STAR:2023pal,STAR:2023ksv}) attempts to eliminate or reduce the impact of selection bias on the studies of nuclear modifications. 
Some experimentalists have suggested that the issue of selection bias could be resolved by canceling the transverse momentum cut on the jet spectrum~\cite{ALICE:2023qve,ALICE:2023jye,STAR:2023pal,STAR:2023ksv}.
In addition, one can efficiently track the jet evolution in the Monte Carlo simulations for the theoretical models. In
this work, we study the influence of selection bias on our concerned observable from the effect of
jet modification using a ``matching-jet'' method, first proposed in
Refs.~\cite{Brewer:2020chg,Brewer:2021hmh}. 
Specifically, in the two hadron-level inclusive jet samples, quenched and unquenched, we pair two jets whose distance in the $(\eta, \phi)$ plane is less than the jet radius $R$ before and after quenching, considering them as the unquenched and quenched versions of the same jet.
To minimize
the omission of jets that have experienced sizeable energy loss, we have appropriately reduced the
transverse momentum threshold in the reconstruction of quenched jets ($\pT>10~\GeV$) to take into
account the region of I and II in Figure~\ref{fig:pt_selection_bias}. Since the contribution of
region III is small, and the matching efficiency in this region is poor, we neglect this region in
the matching procedure. The matching procedure lets us focus on the jet-by-jet comparison before
and after the in-medium evolution. Although this method makes it difficult to consider the effects
of the nuclear parton distribution function and could not be applied to experimental measurements,
it gives us a new perspective to study the medium modification of the angle between the Standard
and WTA jet axes.

\begin{figure}
	\centering
	\includegraphics[width=1\columnwidth]{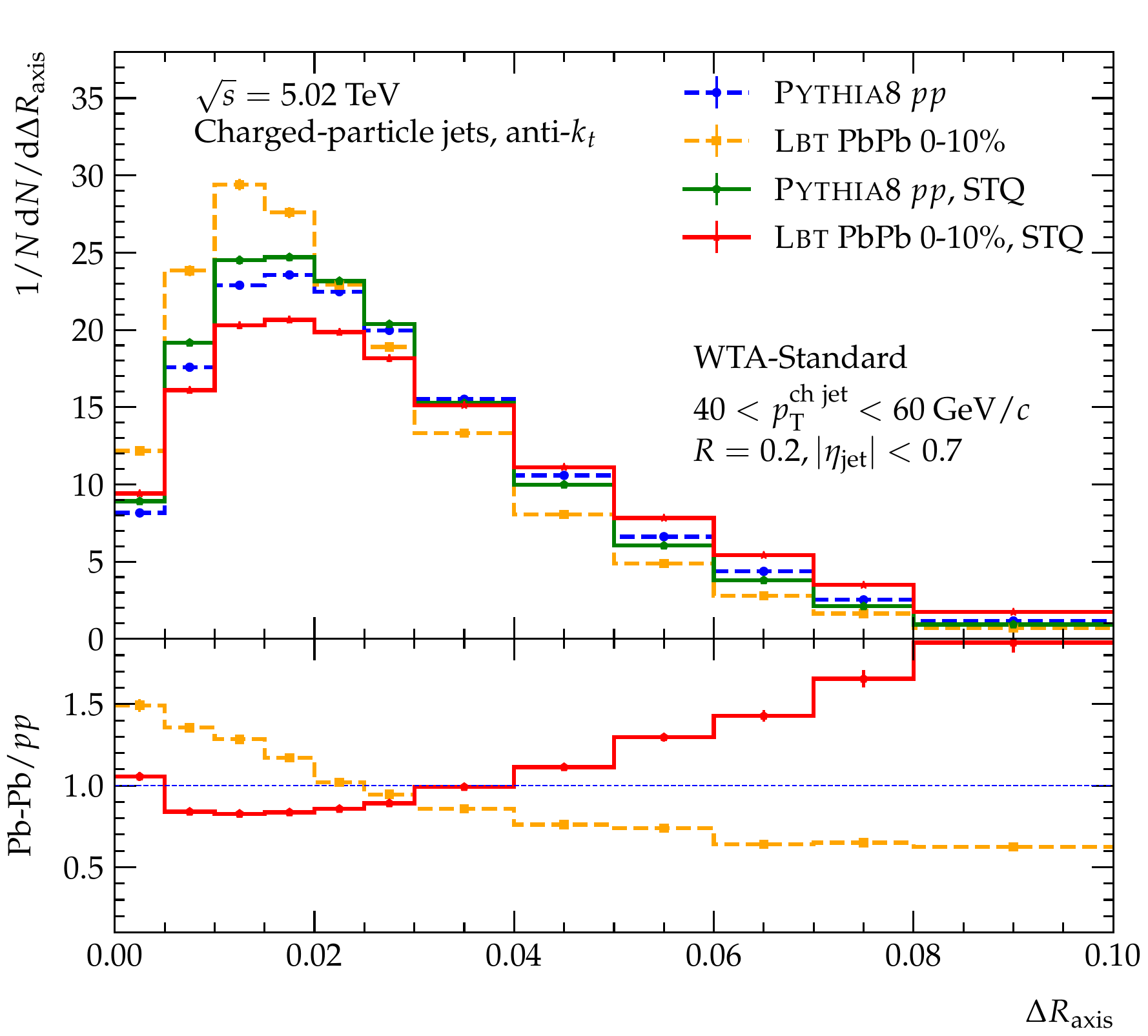}
	\caption{Normalized \DeltaRWtaStd{} distributions of the selected jet samples in \pp\ and
		0-10\% \PbPb\ collisions at $\sqrt{s_{NN}}=5.02~\TeV$. STQ represents the calculations
		with matching-jet method. The ratios are also plotted in the lower panel.
	}
	\label{fig:STQ}
\end{figure}

In Figure~\ref{fig:STQ}, we show the distributions of \DeltaRWtaStd{} for samples of matched jets
in the region I$+$II. For comparison, the Monte Carlo simulation results shown in
Figure~\ref{fig:model_vs_experimental} have also been added to this figure. For the convenience of
the following discussion, we labeled these matched jets by Select-Then-Quench (STQ), a matching-jet
method first introduced in Ref.~\cite{Brewer:2020chg}. Since the jets are well-tracked in the
matching procedure, comparing the STQ ($pp$) and STQ (PbPb) cases reveals the actual jet
modification in the QGP, in which the influences of selection bias are almost removed. 
We observe that the \DeltaRWtaStd{} distribution of STQ (PbPb) shows evident broadening compared to that of STQ
($pp$), and the PbPb/$pp$ ratio of STQ is enhanced at $\DeltaR>0.04$. 
It means the jets get
broader by the interactions with the QGP medium compared to their initial structures. However, the
modification patterns of STQ results are inverse to the normal theoretical calculation, while the
latter contains the effect of selection bias. 
Now, we can conclude that the selection bias obscures the real nature of intra-jet broadening and leads to a narrowed modification of the \DeltaRWtaStd{} distribution.
Many researchers have been focused on searching for medium-induced intra-jet broadening~\cite{Wang:2013cia,Ringer:2019rfk}, and recent experimentalists have also reported some positive signals~\cite{STAR:2023pal,STAR:2023ksv,ALICE:2023jye}.

\begin{figure}
	\centering
	\includegraphics[width=1\columnwidth]{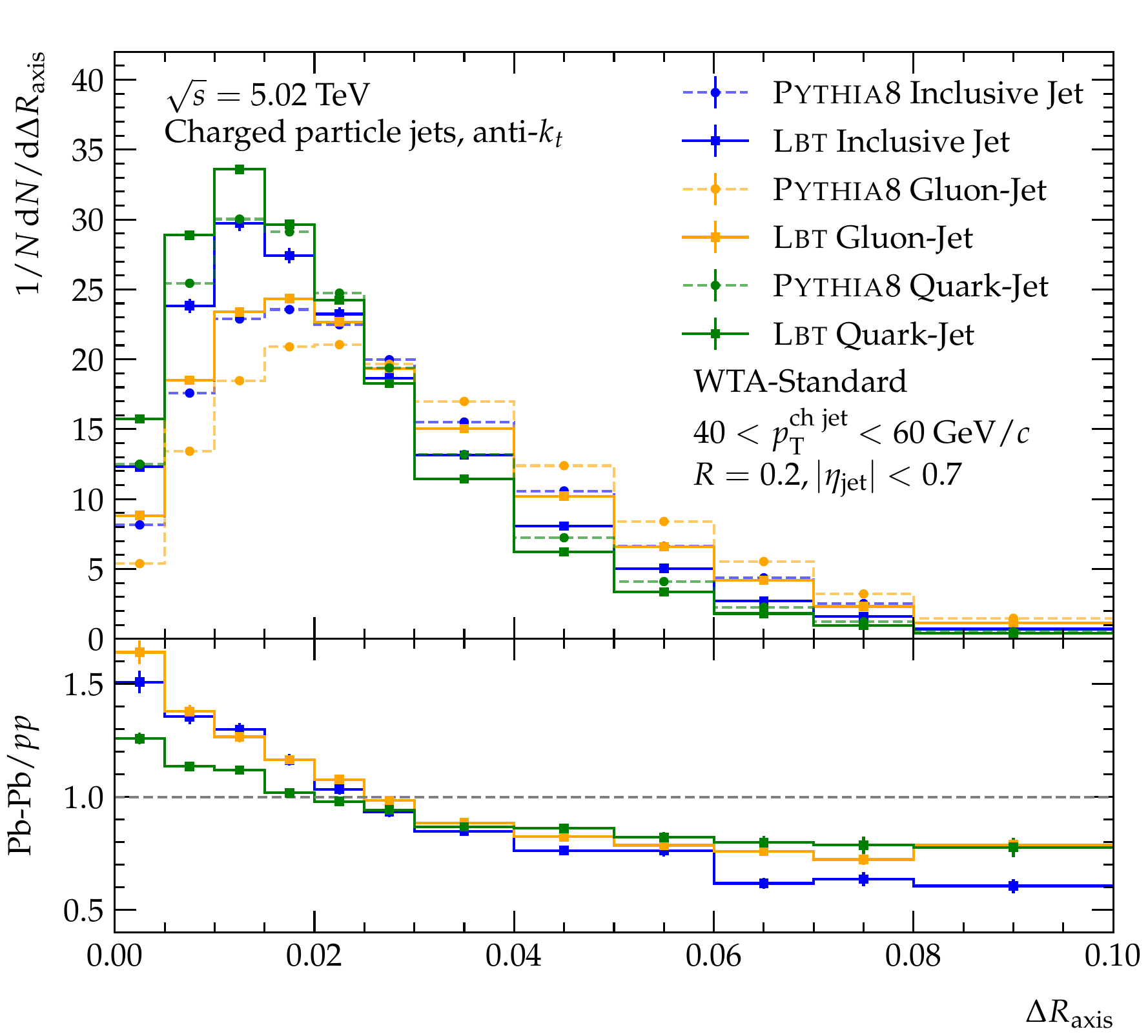}
	\caption{Normalized \DeltaRWtaStd{} distributions of the quark-jets, gluon-jets and
		inclusive jets in \pp\ and 0-10\% \PbPb\ collisions at $\sqrt{s_{NN}}=5.02~\TeV$.
		The ratios are also plotted in the lower panel.
	}
	\label{fig:Quark Gluon Jets dR}
\end{figure}

This simple test above demonstrates the importance of selection bias in studying the nuclear
modification of the \DeltaR\ distribution. However, please note that although we have excluded
selection bias using the STQ method, this also obscures the quark-initiated jets' fraction increase effect
that existed in experiments or conventional Monte Carlo studies, where the proportion of
quark-initiated jets tends to increase because gluons usually lose more energy~\cite{Yan:2020zrz,Qiu:2019sfj,Qiu:2020pus}. In Monte Carlo
studies, we can distinguish whether the jet is a quark- or gluon-jet by comparing the distance
between the jet and the initial outgoing hard partons in the $(\eta, \phi)$ plane. Utilizing this
method, we have employed the \PYTHIA{8} event generator to ascertain that the quark- and gluon-jet
fractions of inclusive jets in \pp{} collisions at \snn{5.02}{TeV}\ at leading-order (LO) accuracy
are about 36.7\% and 63.3\% with jet radius $R=0.2$, $|\eta_{\mathrm{jet}}|<0.7$ and
$40<\pT^{\mathrm{ch~jet}}<60~\GeVc$. To test the flavor dependence of jet \DeltaR{} and its medium
modification, as shown in Figure~\ref{fig:Quark Gluon Jets dR}, we estimate the \DeltaR{}
distributions of quark-jets, gluon-jets and inclusive jets in \pp\ and 0-10\% \PbPb\ collisions at
$\sqrt{s_{NN}}=5.02~\TeV$. The quark-jets exhibit a narrower initial \DeltaR{} distribution than
the gluon-jets. In \PbPb\ collisions, we find that both the quark- and gluon-jets become narrower
than in \pp, and the narrowing of the latter is more substantial. Moreover, the \DeltaR{}
modification of inclusive jets seems closer to that of the gluon-jets sample.

\begin{figure}
	\centering
	\includegraphics[width=1\columnwidth]{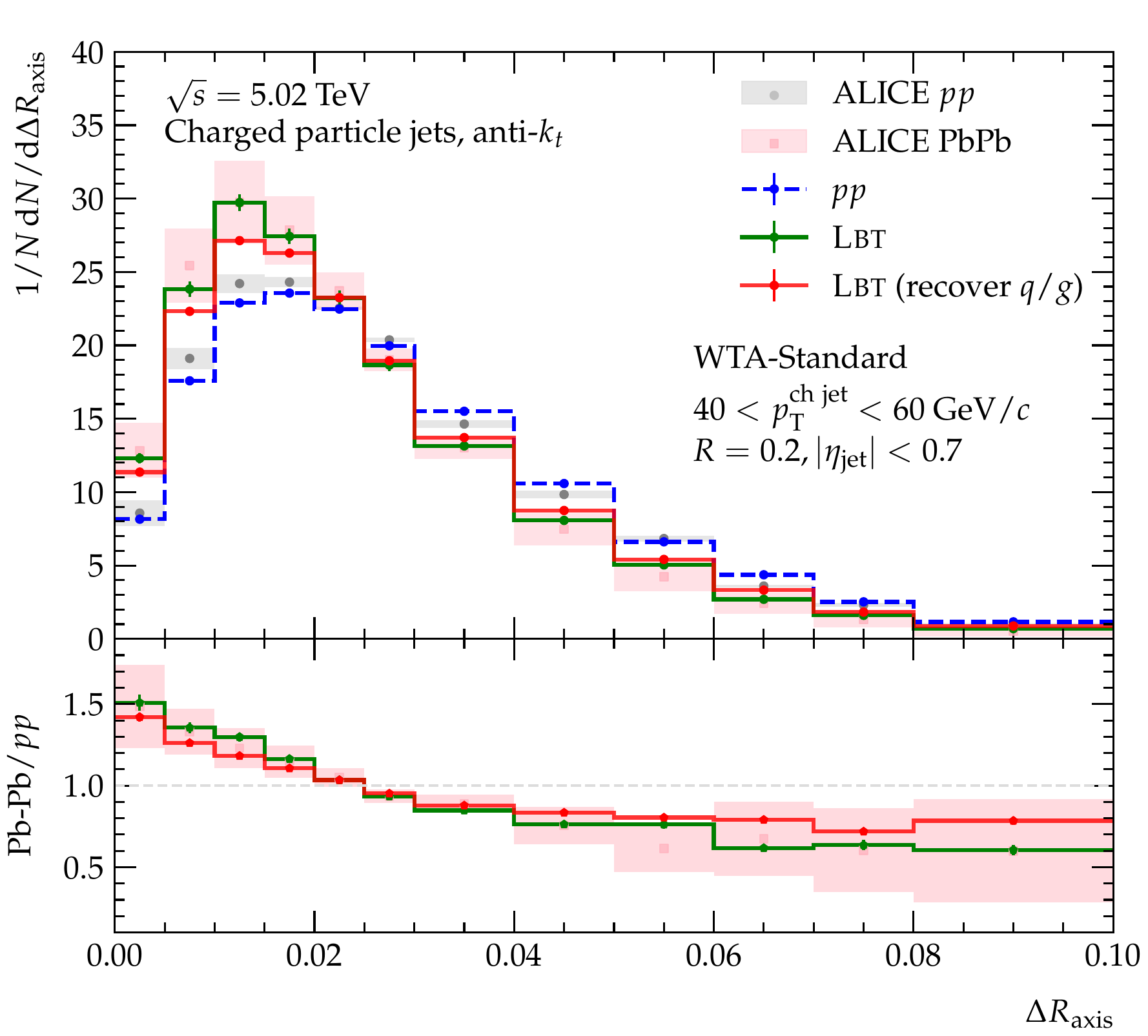}
	\caption{The LBT calculations of \DeltaRWtaStd{} distributions in Pb+Pb collisions with
		recovered quark/gluon-jet fractions are compared with the normal LBT calculations and
		the ALICE data. The ratios are also plotted in the lower panel.}
	\label{fig:QG fraction}
\end{figure}

To test the influence of the quark/gluon-jet fraction change on the \DeltaRWtaStd{} modification,
we rescale their fractions in \PbPb\ collisions to be consistent with their initial fraction
(36.7\% and 63.3\%). The calculations with rescaled quark/gluon-jet fractions, denoted as \LBT\
(recover $q/g$), are shown in Figure~\ref{fig:QG fraction} and compared to the normal \LBT\
calculations as well as the ALICE data. We find that the calculations with rescaled fractions do
not significantly differ from the normal \LBT\ calculations. Since the rescaled calculations
exclude the influence of the fraction changes, it hints that the decreased gluon-jet fraction is
not the main reason that leads to a narrowing \DeltaR{} distribution of inclusive jet in \PbPb\
collisions.

\begin{figure}
	\centering
	\includegraphics[width=1\columnwidth]{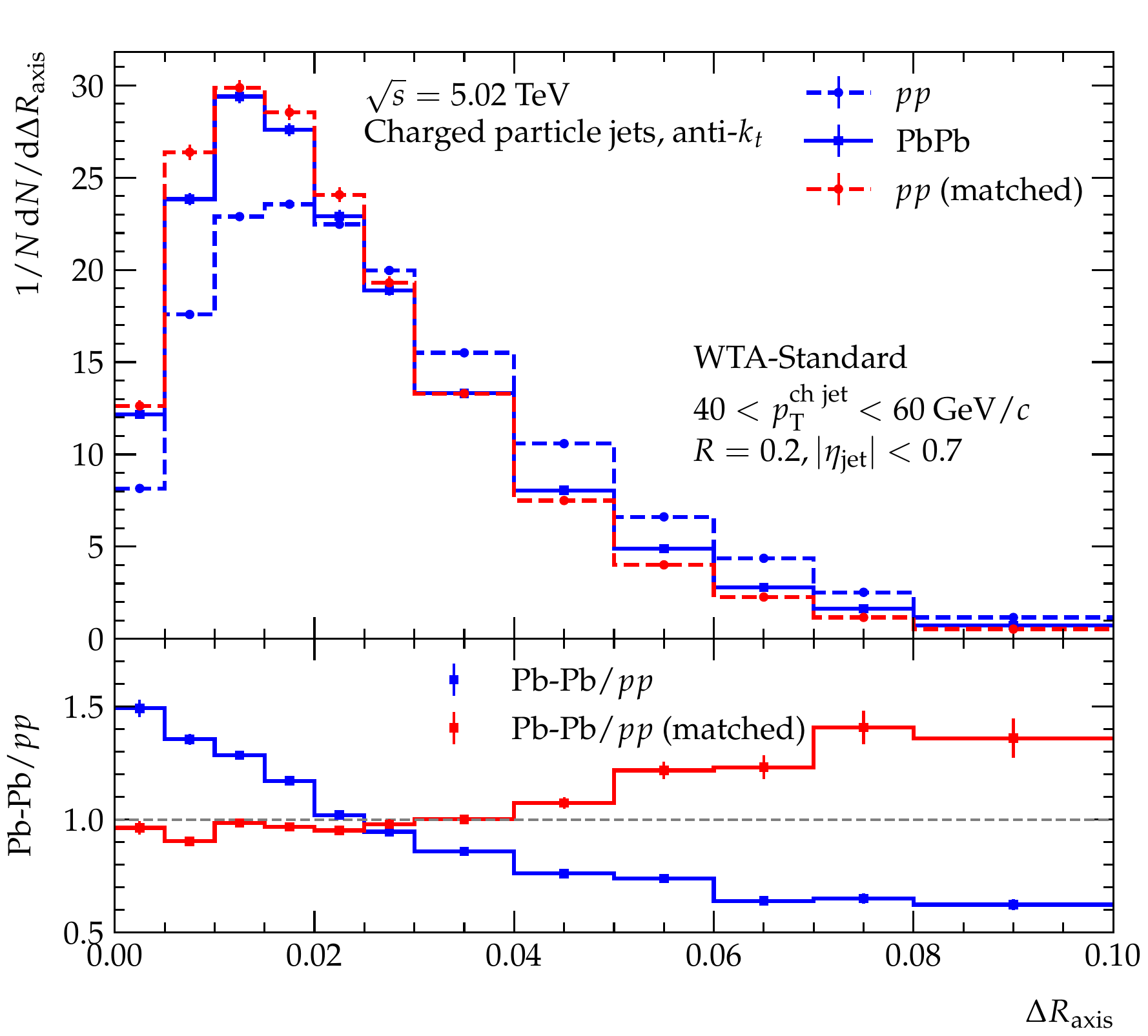}
	\caption{
		The \DeltaRWtaStd{} distribution of the selected jet sample with $40<\pT<60~\GeV$
		in \PbPb\ is compared to its initial counterpart, denoted as $pp$(matched), and
		that selected in \pp\ collisions, the ratios of PbPb/$pp$ and PbPb/$pp$(matched)
		are also plotted in the lower panel.
	}
	\label{fig:QTS}
\end{figure}

To find out the actual jet modification, in Figure~\ref{fig:STQ}, we have compared the selected jet
sample with $40<\pT<60~\GeV$ in \pp\ collisions with its quenched counterpart in \PbPb\ collisions
using the matching-jet method. Similarly, we can also trace the initial counterpart of the jet
sample with $40<\pT<60~\GeV$ selected in \PbPb\ collisions. We denote such matched initial
counterpart as $pp$(matched) for simplicity. In Figure~\ref{fig:QTS}, the \DeltaRWtaStd{}
distribution of the selected jet sample with $40<\pT<60~\GeV$ in \PbPb\ is compared to its initial
counterpart, the $pp$(matched), and the one with $40<\pT<60~\GeV$ selected in \pp\ collisions; the
ratios of PbPb/$pp$ and PbPb/$pp$(matched) are also plotted in the lower panel. We find that
$pp$(matched) has a significantly narrower distribution than the \pp. The jet sample selected in
\AAcoll\ collisions after jet quenching usually has higher initial \pT, while the one with higher
\pT{} usually has narrower \DeltaRWtaStd{} as shown in Figure~\ref{fig:pp-baseline}. When taking the
ratio of \PbPb\ to their initial counterpart, we can observe a broadening in \DeltaRWtaStd{} as
shown in the lower panel of Figure~\ref{fig:QTS}. However, if we only compare the jet samples with
$40<\pT<60~\GeV$ selected independently in \pp\ and \PbPb\ collisions, the broadening turns into
narrowing. In other words, when taking the ratio of PbPb/$pp$ in the ALICE measurements, the
selected jet sample in A+A collisions has a significantly narrower initial \DeltaRWtaStd{}
distribution than the \pp\ baseline. 
Such a biased comparison between \pp\ and A+A might conceal the real intra-jet broadening behavior during the jet quenching.

Finally, another point to note is that our work has not considered nPDFs effects in the \PbPb\ collision simulations. 
Our study investigated how cold nuclear effects affect \DeltaRWtaStd\ distributions in \PbPb\ collisions. The overall trend of nuclear modification showed no significant difference whether we included these cold nuclear effects in the model or not. 
To facilitate the application of the ``matching-jet'' method for matching \PbPb\ events with \pp\ references, we deliberately omitted nPDF effects in this work.

\section{Summary}
\label{sec:sum}
In this work, we study the angle between the Standard and Winner-Take-All jet axes
\DeltaRWtaStd{} in \pp\ and 0-10\% \PbPb\ collisions at $\sqrt{s_{NN}}=5.02~\TeV$. We mainly
investigate the influences of selection bias and the quark/gluon-jet fraction changes on the
modification patterns of the \DeltaRWtaStd{} distribution in \PbPb\ collisions. The initial
\DeltaRWtaStd{} distribution in \pp\ collisions is calculated by the \PYTHIA{8} event generator.
The in-medium jet propagation in nucleus-nucleus collisions is simulated by the \LBT\ model, which
considers both the elastic and inelastic jet-medium interactions.

Firstly, our theoretical results calculated by the \LBT\ model show that the \DeltaRWtaStd{}
distribution in \PbPb\ at \snn{5.02}{TeV}\ is narrower than that in \pp, which agrees well with the
recently reported ALICE measurements. However, the narrowing of \DeltaRWtaStd{} seems to violate the nature of intra-jet broadening due to jet quenching. 
The phenomenon may be attributed to ``selection bias'', which skews the jet selection process by preferentially including candidates that exhibit less energy loss in \AAcoll\ collisions.
Utilizing a matching-jet method to track the jet evolution in the
QGP to reduce the selection bias in the Monte Carlo simulations, we observe that the
\DeltaRWtaStd{} distribution becomes broader due to the jet-medium interactions. At the same time,
by rescaling the quark/gluon-jet fractions in \PbPb\ collisions to be the same as that in \pp, we
find that the fraction change may not significantly influence the modification pattern of jet
\DeltaRWtaStd{}. On the other hand, we find that the selected jet sample in A+A collisions has a
significantly narrower initial \DeltaRWtaStd{} distribution than the \pp\ baseline, and such a
biased comparison between \pp\ and A+A might conceal the actual intra-jet broadening effect in the
experimental measurements. 
The investigations presented in this paper will deepen our understanding
of the relationship between the actual intra-jet modifications in the QGP and the experimental
observations.
However, the methodology currently utilized in this study cannot be directly applied to experimental measurements.
Recently, we have noticed that Z/$\gamma$-tagged jets or high-\pT-hadron-tagged jets can be used to reduce selection bias and can apply to real experimental measurements~\cite{ALICE:2023jye,STAR:2023ksv,STAR:2023pal,CMS:2024zjn,Wang:2024plm}. 
Specifically, by using tagger particles rather than jets for event selection, the selection bias is only induced by the choice of tagger. Since tagger particles do not interact with the QGP medium,
this approach significantly reduces the impact of selection bias on jet quenching studies.
Our future research will employ Z/$\gamma$-tagged jets to minimize selection bias and explore intra-jet broadening phenomena in heavy-ion collisions.

\paragraph*{Acknowledgments:} 
This research is supported by the Guangdong Major Project of Basic and
Applied Basic Research No. 2020B0301030008, and the National Natural Science Foundation of China
with Project Nos. 11935007, 12035007 and 12247127. S. Wang is supported by China Postdoctoral
Science Foundation under project No. 2021M701279.

\bibliographystyle{apsrev4-1}
\bibliography{refs}
\end{document}